\documentclass[aps,prl,twocolumn,superscriptaddress,longbibliography]{revtex4}
\usepackage{graphicx}
\usepackage{amssymb,amsmath}
\usepackage{bm}
\usepackage{dcolumn}
\usepackage{subfigure}
\usepackage{float}
\usepackage[OT1]{fontenc} 
\usepackage{url}
\usepackage{slashed}
\usepackage{color}
\usepackage{verbatim}
\usepackage{times}
\usepackage{graphicx}
\usepackage{hyperref}

\begin{document}
	
\title{Expressivity of Quantum Neural Networks}

\author{Yadong Wu}
\affiliation{Institute for Advanced Study, Tsinghua University, Beijing, 100084, China}

\author{Juan Yao}
\affiliation{Guangdong Provincial Key Laboratory of Quantum Science and Engineering, Shenzhen Institute for Quantum Science and Engineering, Southern University of Science and Technology, Shenzhen 518055, Guangdong, China}

\author{Pengfei Zhang}
\email{pengfeizhang.physics@gmail.com}
\affiliation{Institute for Quantum Information and Matter, California Institute of Technology, Pasadena, California 91125, USA}
\affiliation{Walter Burke Institute for Theoretical Physics, California Institute of Technology, Pasadena, California 91125, USA}

\author{Hui Zhai}
\email{hzhai@tsinghua.edu.cn}
\affiliation{Institute for Advanced Study, Tsinghua University, Beijing, 100084, China}

\begin{abstract}

In this work, we address the question whether a sufficiently deep quantum neural network can approximate a target function as accurate as possible. 
We start with simple but typical physical situations that the target functions are physical observables, and then we extend our discussion to situations that the learning targets are not directly physical observables, but can be expressed as physical observables in an enlarged Hilbert space with multiple replicas, such as the Loshimidt echo and the Renyi entropy. The main finding is that an accurate approximation is possible only when the input wave functions in the dataset do not exhaust the entire Hilbert space that the quantum circuit acts on, and more precisely, the Hilbert space dimension of the former has to be less than half of the Hilbert space dimension of the latter. In some cases, this requirement can be satisfied automatically because of the intrinsic properties of the dataset, for instance, when the input wave function has to be symmetric between different replicas. And if this requirement cannot be satisfied by the dataset, we show 
that the expressivity capabilities can be restored by adding one ancillary qubit where the wave function is always fixed at input. Our studies point toward establishing a quantum neural network analogy of the universal approximation theorem that lays the foundation for expressivity of classical neural networks.

\end{abstract}
	
\maketitle

Neural networks lie at the center of the recent third trend of artificial intelligence. The universal approximation theorem plays an essential role in the developments of neural networks, which states that sufficiently wide or sufficiently deep neural networks can approximate a well-behaved function on $d$-dimensional Euclidean space ${\bf R}^d$ with arbitrary accuracy. This theorem lays the foundation of the expressive capability of neural networks and serves as bases for the successes of neural network applications. Quantum neural networks (QNN) are quantum generalizations of classical feedforward neural networks on future quantum computers, which lie at the center of the recent development of the quantum machine learning. However, the expressivity of the QNN has yet been fully explored. 

Here we consider the quantum generalizations of fully connected neural networks, which contain quantum wave functions of $n$-qubit states as inputs, parameterized quantum circuits made of local quantum gates, and measurements on readout qubits leading to labels. The parameters in the quantum circuit will be optimized during training that yields the best approximation of the learning target. Since the Hilbert space dimension of an $n$-qubit state is $2^{n}$, the wave function can encode information of $d=2^{n}$ complex numbers, up to a normalization condition and a global phase. In the following, concerning the effect of depth and width on the expressive capability of a QNN, we address the question whether a sufficiently deep QNN can express any well-behaved function in the ${\bf C}^d$ space

To be concrete, here we consider a number of typical learning tasks in quantum physics problem, where the learning targets include (i) physical observables; (ii) the Loshmidit echo and (iii) the Renyi entropy. We point out that, in contrast to the universal approximation theorem for classical neural networks, a QNN \textit{cannot} express a general well-behaved function with arbitrary accuracy even though the QNN is made sufficiently deep. However, we show in this work that this problem can be solved and the expressivity can be significantly improved, and can even be made as accurate as possible by enlarging the Hilbert space dimension of the input state, which effectively increases the width of the QNN. This can be achieved either by adding an ancillary qubit in the input and (or) by duplicating replicas of the input wave functions. This result points toward an analogy of the universal approximation theorem for QNN. 

\begin{figure}[t]
	\includegraphics[width=.95\columnwidth]{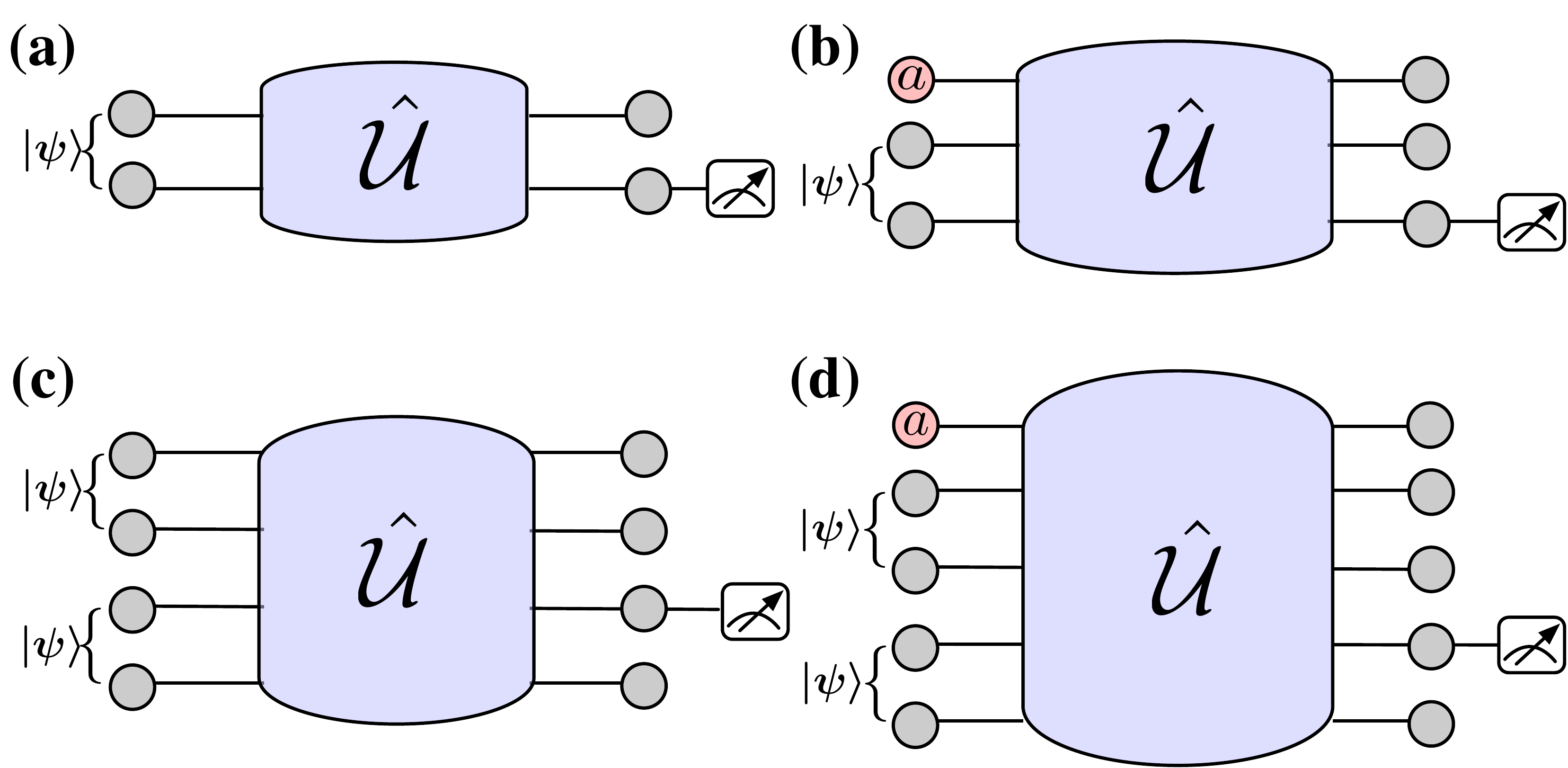}
	\caption{(a) Conventional QNN structure. (b) QNN with one ancillary qubit. (c) QNN with two duplicated replicas of inputs. (d) QNN with two duplicated replicas of input and one ancillary qubit. Here $|\psi\rangle$ are input wave function in the dataset, $\hat{\mathcal{U}}$ denotes the unitary rotation by quantum circuit, and the detector denotes the readout qubit. The red qubit with label $a$ denotes the ancillary qubit at which the wave function is always fixed.} 
	\label{circuit}
\end{figure}

\textbf{Results.} We consider a dataset denoted as $\{(|\psi^l\rangle, y^l)\},\{l=1,\dots,N_\text{D}\}$, where $l$ labels data and $N_\text{D}$ is the total number of data in the dataset. Each input quantum state $|\psi^l\rangle$ can be written as $|\psi^l\rangle=\sum_{m}c^l_{m}|m \rangle$, where $\{|m\rangle\}$ is a complete set of $2^n$ bases of the $n$-qubit Hilbert space, and $\{c^l_{m}\}$ are $2^n$ normalized complex numbers with a fixed total phase. Usually the information in the label is much more condensed than the information of the entire input, therefore, here we consider that the label is simply a number $y^{l}\in [-1,1]$. To motivate this result, let us first start with a simpler situation that the label is a physical observable $y^l=\langle \psi^l|\hat{O}|\psi^l\rangle$, where $\hat{O}$ is a hermitian operator on $n$-qubit quantum state \cite{footnote}. This is equivalent to say that $y^l$ is a quadratic function of these complex numbers as $y^l=\sum_{mm'}O_{mm'} c^{l*}_mc^l_{m'}$, where $O_{mm'}=O^*_{mm'}=\langle m|\hat{O}|m'\rangle$. The QNN we considered is shown in Fig.~\ref{circuit}(a). A unitary $\hat{\mathcal{U}}$ made of local quantum gates acts on the input wave function $|\psi^l\rangle$, and then we perform a measurement, say $\sigma_x$, on a readout qubit-r. The measurement operator is therefore denoted by
\begin{align}
\hat{M}=\hat{\sigma}_0^1\otimes\cdots\otimes\hat{\sigma}_x^r\otimes\cdots\otimes\hat{\sigma}_0^N,
\end{align}  
where the superscription $i=1,\dots,N$ denotes the qubits, and $\sigma^i_0$ denotes the identity matrix. The measurement of the quantum circuits leads to
\begin{equation}
\tilde{y}^l=\langle \psi^l|\hat{\mathcal{U}}^\dag\hat{M}\hat{\mathcal{U}}|\psi^l\rangle. \label{tildey}
\end{equation}
The loss function is taken as $\mathcal{L}=\frac{1}{N_D}\sum_{l}|\tilde{y}^l-y^l|^2$, which enforces $\tilde{y}^l$ to be $y^l$ for all $|\psi^l\rangle$. 

Here one thing that should be noticed is whether $N_\text{D}>2^{n}$ or whether the input wave functions exhaust the entire Hilbert space. If $N_\text{D}<2^n$, all the input wave functions in the dataset only occupy a subset of the entire Hilbert space. In some cases, even when $N_\text{D}>2^{n}$, if the wave functions have certain structures, for instance, if the wave functions are taken as ground states of certain Hamiltonians \cite{QNN17,QNN18}, they also do not exhaust the entire Hilbert space. However, if $N_\text{D}>2^{n}$ and the input wave functions are general enough, they exhaust the entire Hilbert space. In this case, in order for all $\tilde{y}^l$ to faithfully represent $y^l$, one requires $\hat{O}=\hat{\mathcal{U}}^\dag\hat{M}\hat{\mathcal{U}}$. However, this is not possible for a general operator $\hat{O}$. This is because the eigenvalues of $\hat{M}$ consist $2^{n-1}$ number of $-1$ and equal number of $+1$, and any unitary transformation keeps these eigenvalues invariant. That is to say, even though one can make the QNN deep enough to present a generic unitary $\hat{\mathcal{U}}$ in the SU$(2^n)$ group, it always cannot satisfy $\hat{O}=\hat{\mathcal{U}}^\dag\hat{M}\hat{\mathcal{U}}$. This argument can be easily generalized to situations that measurements are performed in more than one readout qubits.   

\textit{Ancillary Qubit.} Now we show this problem can be solved by adding one ancillary qubit. Instead of $|\psi^l\rangle$, we now add one ancillary qubit and the input wave function is set as $|\alpha\rangle \otimes |\psi^l\rangle$, where the input state at the ancillary qubit is always fixed as $|\alpha\rangle$. The unitary $\hat{\mathcal{U}}$ now acts on the entire $2^{n+1}$-dimensional Hilbert space, and the measurement is still performed in the readout qubit and now
\begin{align}
\hat{\tilde{M}}=\hat{\sigma}_0^a\otimes \hat{\sigma}_0^1\otimes\cdots\otimes\hat{\sigma}_x^r\otimes\cdots\otimes\hat{\sigma}_0^N,
\end{align}  
where the superscript $a$ denotes the ancillary qubit. The structure is shown in Fig. \ref{circuit}(b). Now we will show that for any given $|\alpha\rangle$, we can always construct an operator $\hat{\tilde{O}}$, acting on the $2^{n+1}$ Hilbert space, which satisfies the following two requirements. The first is that the operator $\hat{\tilde{O}}$ can generate the observables as $\langle\alpha|\otimes\langle\psi^l\big|\hat{\tilde{O}}\big|\alpha\rangle\otimes|\psi^l\rangle=\langle\psi^l|\hat{O}|\psi^l\rangle=y^l$, and the second is that the eigenvalues of the operator $\tilde{O}$  consist of $2^n$ number of $+1$ and equal number of $-1$ and are consistent to that of the measurement operator $\hat{\tilde{M}}$. Without loss of generality, we choose $|\alpha\rangle=|\uparrow\rangle$, it can be shown that $\hat{\tilde{O}}$ chosen as $\hat{\sigma}_z^a\otimes\hat{O}+\hat{\sigma}_x^a\otimes\sqrt{I-\hat{O}^2}$ satisfies these two conditions. First, 
\begin{align}
 &\langle\uparrow|\otimes\langle\psi^l\big|\hat{\tilde{O}}\big|\uparrow\rangle\otimes|\psi^l\rangle\nonumber\\
 =&\langle\uparrow|\hat{\sigma}_z^a|\uparrow\rangle\langle\psi^l|\hat{O}|\psi^l\rangle+\langle\uparrow|\hat{\sigma}_x^a|\uparrow\rangle\langle\psi^l|\sqrt{I-\hat{O}^2}|\psi^l\rangle \nonumber\\
 =&\langle\psi^l|\hat{O}|\psi^l\rangle=y^l. \label{equal}
\end{align}
Secondly, suppose $\{|m\rangle\}$ is a set of eigenbases in the $2^n$-dimensional Hilbert space (without the ancillary qubit) that $\hat{O}|m\rangle=O_m|m\rangle$, and under these bases, $\hat{\tilde{O}}$ can be written as $\sum_m\left(\hat{\sigma}^a_z O_m+\hat{\sigma}^a_x\sqrt{1-O^2_m}\right)|m\rangle\langle m|$. Therefore, its eigenvalue consist $2^n$ number of $+1$ and equal number of $-1$, which equal the eigenvalues of operator $\hat{\tilde{M}}$. When such an operator $\hat{\tilde{O}}$ is found, it is possible to find a unitary $\hat{\mathcal{U}}$ in the $2^{n+1}$-dimensional space, such that $\hat{\mathcal{U}}^\dag\hat{\tilde{M}}\hat{\mathcal{U}}=\hat{\tilde{O}}$, and then,  
\begin{equation}
\langle\alpha|\otimes\langle\psi^l\big|\hat{\mathcal{U}}^\dag\hat{\tilde{M}}\hat{\mathcal{U}}\big|\alpha\rangle\otimes|\psi^l\rangle=\langle\alpha|\otimes\langle\psi^l\big|\hat{\tilde{O}}\big|\alpha\rangle\otimes|\psi^l\rangle=y^l.
\end{equation}
This shows, with the help of one ancillary qubit, the QNN can accurately express the functional mapping $y^l=\langle \psi^l|\hat{O}|\psi^l\rangle$ for all generic quantum states $|\psi^l\rangle$. 

\begin{widetext}

\begin{figure}[t]
	\includegraphics[width=1\columnwidth]{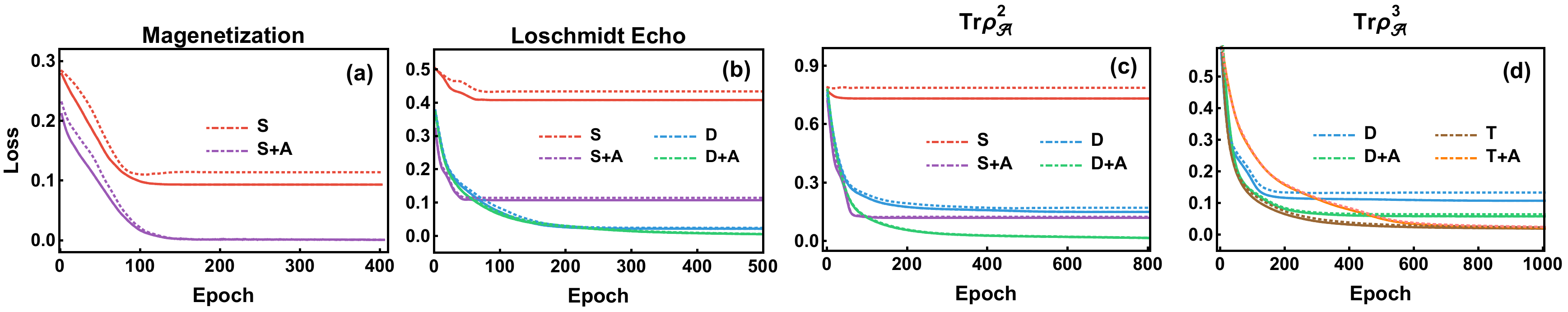}
	\caption{The loss function as a function of the training epoch. The solid lines and the dashed lines are the loss on the training dataset and the validation dataset, respectively. The target functions are respectively the magnetization (a), the Loshmidit echo (b), the second Renyi entropy (c) and the third Renyi entropy (d). For the input wave function $|\psi^l\rangle$ in the dataset, the number of qubit $n=3$ for (a), and $n=2$ for (b), (c) and (d). For the Renyi entropy discussed in (c) and (d), one qubit is taken as $\mathcal{A}$ and the other qubits are  taken as $\mathcal{B}$. In the legends, ${\rm S}$, ${\rm D}$ and ${\rm T}$ respectively denote single, double and triple replicas as input for quantum circuit, and $+{\rm A}$ means that one ancillary qubit is added.  
	\label{loss}}
\end{figure} 

\end{widetext}

In Fig. \ref{loss}(a), we show the loss function for learning total magnetization of generic wave functions in a three-qubit quantum state, with $\hat{O}$ chosen as $\sum_{i}\sigma^{i}_z$. The red lines and the purple lines are results for QNN with structures shown in Fig. \ref{circuit}(a) and (b), respectively. Structure shown in Fig. \ref{circuit}(b) has one more ancillary qubit compared with the structure shown in Fig. \ref{circuit}(a). One can see that, if without the ancillary qubit, the loss clearly saturates to a finite value even for sufficient large training epochs. By adding the ancillary qubit, the loss is significantly reduced and approaches to zero.  

\textit{Lesson.} The lesson from above example is that the learning accuracy can be significantly improved by enlarging the Hilbert space dimension of the input for quantum circuit. Here we present and prove a general statement. Suppose $\mathcal{H}$ is the total Hilbert space of input for quantum circuit, $\text{Dim}(\mathcal{H})$ denotes its Hilbert space dimension. Let us consider $\mathcal{H}=\mathcal{H}_0\bigoplus\mathcal{H}_1$, and suppose all input wave functions in the dataset only reside in $\mathcal{H}_0$. The statement is that, when $\text{Dim}(\mathcal{H}_1)\geqslant\text{Dim}(\mathcal{H}_0)$, we can always find an operator $\hat{\tilde{O}}$ acting on entire Hilbert space $\mathcal{H}$, such that, (i) for any wave function $|\psi\rangle$ in $\mathcal{H}_0$, $\langle \psi |\hat{\tilde{O}}|\psi\rangle=\langle\psi|\hat{O}|\psi\rangle$, 
(ii) the eigenvalues of $\hat{\tilde{O}}$ consist equal number of $+1$ and $-1$, which are the same as the eigenvalue of the measurement operator. Then, it is possible to find a proper $\hat{\mathcal{U}}$ such that $\hat{\tilde{O}}=\hat{\mathcal{U}}^\dag\hat{\tilde{M}}\hat{\mathcal{U}}$.

The construction of $\hat{\tilde{O}}$ is quite similar as the ancillary qubit example. Let us first consider the situation $\text{Dim}(\mathcal{H}_1)=\text{Dim}(\mathcal{H}_0)$, and suppose $\{|m\rangle\}$ is a set of eigenbases in $\mathcal{H}_0$ with $\hat{O}|m\rangle=O_m|m\rangle$. We can then define a set of bases $\{ |\bar{m}\rangle\}$ in $\mathcal{H}_1$ that have one-to-one correspondence with the bases in set $\{|m\rangle\}$, say, each $|\bar{m}\rangle$ corresponds to a $|m\rangle$. Then, we can construct $\hat{\tilde{O}}$ as
\begin{equation}
\begin{aligned}
\hat{\tilde{O}}=&\sum_{m}O_m \big(|m\rangle\langle m|-|\bar{m}\rangle\langle \bar{m}|\big)\\
&+\sqrt{1-O^2_m}\big(|m\rangle\langle \bar{m}|+| \bar{m}\rangle\langle m|\big). \label{tildeO}
\end{aligned}\end{equation}
It is easy to see that $\hat{\tilde{O}}$ constructed as Eq. \ref{tildeO} satisfies the above two requirements. This can be extended to situations $\text{Dim}(\mathcal{H}_1)>\text{Dim}(\mathcal{H}_0)$. In this case, $\mathcal{H}_1$ is larger than the space spanned by $\{|\bar{m}\rangle\}$, and we choose $\hat{\tilde{O}}$ to be diagonal with equal number of $+1$ and $-1$ eigenvalues in the residual Hilbert space. The ancillary qubit is a specific example of this general statement, where $\mathcal{H}_0$ consists states $|\uparrow\rangle\otimes|\psi\rangle$ and $\mathcal{H}_1$ consists states $|\downarrow\rangle\otimes|\psi\rangle$, where $|\psi\rangle$ denotes the input wave functions in the dataset.  

\textit{Replica.} Now we move to consider the learning tasks such as the Loshmidit echo and the Renyi entropy. The Loshmidit echo is an interference between two wave functions, starting from the same input wave function $|\psi^l\rangle$ and evolved by two different Hamiltonians $\hat{H}_a$ and $\hat{H}_b$ for time duration $t$, that is, $y^l=|\langle\psi^l|e^{i\hat{H}_a t}e^{-i\hat{H}_b t}|\psi^l\rangle|^2$. Here we denote $\hat{W}=e^{i\hat{H}_a t}e^{-i\hat{H}_b t}$, and for most Hamiltonians, $\hat{W}$ is a sufficiently chaotic operator for long enough $t$. In Fig. \ref{loss}(b), adopting the QNN in Fig. \ref{circuit} (a) and (b) as before, we show the loss function for learning the Loshmidit echo. We can see that even with an ancillary qubit, the loss still saturates to a finite non-zero value even with sufficient long training epochs. The reason is also quite obvious. It is because for the Loshmidit echo, the label $y$ is a quartic function of $\{c_m\}$, while the QNN yields $\tilde{y}$ given by Eq. \ref{tildey} which is only a quadratic function of  $\{c_m\}$. Thus, to accurately capture learning target such as the Loshmidit echo, non-linearity is necessary. 

There are also various discussions on adding non-linearity in QNN. Here we show that duplicating replica of the input states is another way to incorporate the non-linearity. In fact, it is a quite efficient way in this case, which can be easily seen from 
\begin{align}
y^l&=|\langle\psi^l|\hat{W}|\psi^l\rangle|^2
=\langle\psi^l|\otimes\langle \psi^l|\hat{W}\otimes\hat{W}^\dagger|\psi^l\rangle\otimes|\psi^l\rangle\nonumber\\
&=\langle\psi^l|\otimes\langle \psi^l|\hat{W}^\dag\otimes\hat{W}|\psi^l\rangle\otimes|\psi^l\rangle\nonumber\\
&=\langle\psi^l|\otimes\langle \psi^l|\frac{1}{2}\left(\hat{W}^\dag\otimes\hat{W}+\hat{W}^\dag\otimes\hat{W}\right)|\psi^l\rangle\otimes|\psi^l\rangle.
\end{align} 
Suppose the input wave function is a $n$-qubit state, and when we double the input to a $2n$-qubit state, the Loshmidit echo returns to a quadratic function in the enlarged Hilbert space. In the doubled space, the Loshmidit echo becomes a physical observable with $\hat{O}=(\hat{W}^\dag\otimes\hat{W}+\hat{W}^\dag\otimes\hat{W})/2$ being a Hermitian operator. 

We can also consider another example of the Renyi entropy. For an input wave function $|\psi^l\rangle$, by partially tracing out a sub-system $\mathcal{B}$, the reduced density matrix for remaining sub-system $\mathcal{A}$ is given by $\rho_\mathcal{A}=\text{Tr}_\mathcal{B}|\psi^l\rangle\langle \psi^l|$. Considering the label $y^l$ as the $m$th order Renyi entropy given by $y^l=\text{Tr}\rho^m_\mathcal{A}$, we show in Fig. \ref{loss}(c) the loss function for learning the second Renyi entropy. Similar as the Loshmidit echo case, without replica the loss still saturates to a finite non-zero value at sufficient long training epochs even with an ancillary qubit. Similarly, for the second Renyi entropy, it can be shown that
\begin{align}
y^l=\text{Tr}\rho^2_\mathcal{A}
=\langle\psi^l|\otimes\langle \psi^l|\hat{\mathcal{X}}_\mathcal{A}\otimes\hat{\mathcal{I}}_\mathcal{B}|\psi^l\rangle|\otimes |\psi^l\rangle,
\end{align}
where $\hat{\mathcal{X}}_\mathcal{A}$ is the swap operator of $\mathcal{A}$ subsystem between two replicas, and $\hat{\mathcal{I}}_\mathcal{B}$ is the identity operator. Then, the second Renyi entropy becomes physical observable in doubled Hilbert space. This can also be generalized to higher order Renyi entropy, and in general, for the $m$th order Renyi entropy, 
\begin{align}
y^l=\text{Tr}\rho^m_\mathcal{A}
=\langle\psi^l|^{\otimes m}\hat{\mathcal{X}}_\mathcal{A}\otimes\hat{\mathcal{I}}_\mathcal{B}|\psi^l\rangle^{\otimes m},
\end{align}
for which we need $m$ replicas. 

Hence, we double the size of the input for the quantum circuit and duplicate two replicas of the input wave functions as the input. The unitary $\hat{\mathcal{U}}$ then acts on the total Hilbert space with $\text{Dim}(\mathcal{H})=2^{2n}$, with the same measurement on the readout qubit as discussed above. The structure is shown in Fig. \ref{circuit}(c). Now the question is that, with enlarged Hilbert space, whether we still need an ancillary qubit, as shown in Fig. \ref{circuit}(d). Note that the input wave functions are all subjected to a constraint that they have to be symmetric between two replicas, therefore, these wave functions do not exhaust all $2^{2n}$ dimensional Hilbert space. Let us denote such symmetric Hilbert space as $\mathcal{H}_0$, and with the lesson we discussed above, it is important to analyze whether $\text{Dim}(\mathcal{H}_0)$ is larger than the half of $\text{Dim}(\mathcal{H})$. It can be shown that 
\begin{equation}
\text{Dim}(\mathcal{H}_0)=2^{n}+\frac{2^n(2^n-1)}{2}=2^{2n-1}+2^{n-1},
\end{equation}
and it is larger than $\text{Dim}(\mathcal{H})/2$. In other word, because $\text{Dim}(\mathcal{H}_1)<\text{Dim}(\mathcal{H}_0)$, one still needs the ancillary qubit in order to construct a proper $\hat{\tilde{O}}$. This can be seen from Fig. \ref{loss}(b) and (c) for the cases of learning the Loshmidit echo and the second Renyi entropy, respectively. One can see that, even with doubled input, the loss can still be reduced by adding an ancillary qubit. 

The situation becomes different when one considers tripled Hilbert space, for instance, when considering the third Renyi entropy. For the tripled Hilbert space, if we still require the wave functions to be symmetric between three replicas, the Hilbert space dimension $\text{Dim}(\mathcal{H}_0)$ is given by
\begin{align}
&\text{Dim}(\mathcal{H}_0)=2^{n}+2^n(2^n-1)+\frac{2^n(2^n-1)(2^n-1)}{6}\\
&=\frac{1}{3}\left(2^{3n-1}+3\times 2^{2n-1}+2^{n}\right)\leq\frac{1}{2} 2^{3n}=\frac{1}{2}\text{Dim}(\mathcal{H}).
\end{align}
Therefore, in this case, the requirement for finding a proper $\hat{\tilde{O}}$ can be satisfied without adding an ancila qubit. This can be seen in Fig.~\ref{loss}(d). One can see with two replicas, the loss cannot be reduced to a sufficient small value, for both cases without and with the ancila qubit. However, when there are three replicas, even without an ancila qubit, the loss can already drop to sufficiently close to zero. And for sufficiently long training epoch, the losses for QNN with or without ancila qubit approach the same value. This shows that the ancila qubit is not necessary in this case when there are three replicas. And the same conclusion can be generalized to situations with more than three replicas.  

\textbf{Conclusion and Outlook.} In this work, we consider the expressivity of QNN for learning targets that are observables (i.e. expectations of a hermitian operator) of input wave functions. These also include the situations that the learning targets are not observable of input wave functions, but can be expressed as observables in the enlarged Hilbert space with multiple replicas of input wave functions, such as the Loshimidt echo and the Renyi entropy. The main finding of this work is that such target can be expressed accurately only when the input wave functions in all dataset only occupy a subset $\mathcal{H}_0$ of the entire Hilbert space $\mathcal{H}$ that the quantum circuit acts on, especially, we require the condition $\text{Dim}(\mathcal{H}_0)<\text{Dim}(\mathcal{H})/2$. An accurate approximation of the learning target is possible for a sufficiently deep QNN either when this condition is satisfied naturally by the dataset, or when the condition is enforced by artificially adding an ancillary qubit. 

Our discussions also provide a general recipe for improving learning accuracy, provided that no prior knowledge of such learning task is known. First, one can first try to add an ancillary quibit. If not satisfactory, then one can duplicate two replicas with an ancillary qubit. Finally, if still not satisfied, one can add more replicas, and when the number of replica equals or is greater than three, the ancillary qubit is no longer needed.  

In the future, we can consider a number of generalizations of such studies. First, here we focus on learning targets that are observables or generalized observables, and we can consider more sophisticated learning targets. Secondly, here we focus on regression tasks, and we can consider classification tasks. Thirdly, here we focus on the fully connected architectures, and we can consider other architectures of QNN, such as convolutional QNN and recurrent QNN. We hope such studies can lead to analogy of universal approximation theorem for QNN and lay the foundation of the expressive power of QNN.

\textit{Acknowledgment.} This work is supported by Beijing Outstanding Young Scientist Program, NSFC Grant No. 11734010, MOST under Grant No. 2016YFA0301600. 
J. Y. acknowledge the National Natural Science Foundation of China under Grant No. 11904190. PZ acknowledges support from the Walter Burke Institute for Theoretical Physics at Caltech.


\begin{thebibliography}{99}
\bibitem{footnote}
We can normalize $\hat{O}$ operator such that all its eigenvalues are between $-1$ and $+1$.
\end{thebibliography}
\end{document}